\documentclass[12pt]{extarticle}
\usepackage{setspace}

\usepackage{setspace}
\usepackage[T1]{fontenc}
\usepackage{times}
\usepackage{palatino} 
\usepackage{lmodern}
\usepackage[latin1]{inputenc}
\usepackage{epsfig}
\usepackage[english]{babel}
\usepackage{color}
\usepackage{graphicx}
\usepackage{dcolumn}
\usepackage{moreverb}
\usepackage{amsmath,amssymb,amsfonts}
\usepackage[all]{xy}

\begin{document}
\numberwithin{equation}{section}
\newcommand{\boxedeqn}[1]{%
  \[\fbox{%
      \addtolength{\linewidth}{-2\fboxsep}%
      \addtolength{\linewidth}{-2\fboxrule}%
      \begin{minipage}{\linewidth}%
      \begin{equation}#1\end{equation}%
      \end{minipage}%
    }\]%
}


\newsavebox{\fmbox}
\newenvironment{fmpage}[1]
     {\begin{lrbox}{\fmbox}\begin{minipage}{#1}}
     {\end{minipage}\end{lrbox}\fbox{\usebox{\fmbox}}}

\raggedbottom
\onecolumn

\parindent 8pt
\parskip 10pt
\baselineskip 16pt
\noindent\title*{{\LARGE{\textbf{A new family of $N$ dimensional superintegrable double singular oscillators and quadratic algebra $Q(3)\oplus so(n) \oplus so(N-n)$ }}}}
\newline
\newline
Md Fazlul Hoque, Ian Marquette and Yao-Zhong Zhang
\newline
School of Mathematics and Physics, The University of Queensland, Brisbane, QLD 4072, Australia
\newline
E-mail: m.hoque@uq.edu.au; i.marquette@uq.edu.au; yzz@maths.uq.edu.au
\newline
\newline
\begin{abstract}
We introduce a new family of $N$-dimensional quantum superintegrable model consisting of double singular oscillators of type $(n,N-n)$. The special cases $(2,2)$ and $(4,4)$ were previously identified as the duals of 3- and 5-dimensional deformed Kepler-Coulomb systems with $u(1)$ and $su(2)$ monopoles respectively. The models are multiseparable and their wave functions are obtained in $(n,N-n)$ double-hyperspherical coordinates. We obtain the integrals of motion and construct the finitely generated polynomial algebra that is the direct sum of a quadratic algebra $Q(3)$ involving three generators, $so(n)$, $so(N-n)$ (i.e. $Q(3)\oplus so(n) \oplus so(N-n)$ ). The structure constants of the quadratic algebra themselves involve the Casimir operators of the two Lie algebras $so(n)$ and $so(N-n)$. Moreover, we obtain the finite dimensional unitary representations (unirreps) of the quadratic algebra and present an algebraic derivation of the degenerate energy spectrum of the superintegrable model. 
\end{abstract}
\section{Introduction}

The isotropic and anisotropic harmonic oscillators \cite{Jau1,Bak1,Mos1,Lou1,Bar1,Fra1,Hwa1} are among the most well known maximally superintegrable systems with applications in various areas of physics. However, the aspect of symmetry algebras generated by well-defined integrals of motion is a complicated issue in quantum mechanics as recognized in early work by Jauch and Hill \cite{Jau1}. In the case of the isotropic harmonic oscillator in $N$-dimensional Euclidean space one can apply finite dimensional unitary representations (unirreps) and Gel'fand invariants and their eigenvalues of the Lie algebra $su(n)$ to provide an algebraic derivation  of the spectrum and the degeneracies \cite{Hwa1}. The 3D case of isotropic harmonic oscillator was discussed using what is now called the Fradkin tensor and connected to $su(3)$ generators \cite{Fra1}. There are however various embeddings of the symmetry algebra, some with operators that do not commute with the Hamiltonian \cite{Lou1}, in terms of $su(n+1)$, $su(n,1)$ and $sp(n)$. The anisotropic case has been for a long time the subject of research due to its various applications in nuclear physics. An analysis based on well-defined integrals of motion and polynomial algebras of arbitrary order is known only in the 2D case \cite{Bon1}.\par
%
%
It is also known for a long time that the harmonic oscillator in 2-dimensional Euclidean space is related to the 2D Kepler-Coulomb system  via the so-called Levi-Civita or regularisation transformation \cite{Lev1}. This transformation however can only be extented to certain specific dimensions. The 3- and the 5-dimensional Kepler-Coulomb systems are related to the harmonic oscillators in 4- and 8-dimensional Euclidean space respectively via the Kustaanheimo-Stiefel and the Hurwitz transformations \cite{Kus1,Hur1} and these results can be extended to curved spaces \cite{Kal1}. The Kustaanheimo-Stiefel transformations are connected to the so-called St\"{a}ckel transformations and were used to classify superintegrable systems in conformally flat spaces \cite{mil1,mil2,mil3}. These are specific Levi-Civita, Kustaanheimo-Stiefel and Hurwitz transformations. The above-mentioned connections between various models can be reinterpreted in terms of monopole interactions: the 4D harmonic oscillator has a duality relation with 3-dimensional Kepler system with an Abelian $u(1)$ monopole ( also refered to as MICZ-Kepler systems) \cite{McI1,Zwa1,Jac1,Bar2,Bac1,Dho1,Feh1} and the 8D harmonic oscillator is in fact dual to 5D Kepler-Coulomb system with a non-Abelian $su(2)$ monopole (Yang -Coulomb monopole) \cite{Mard1,Mard2,Mard3,Ner1,Mard4, Ple1,Ple2,Ple3,Tru1}. \par
%
%
It has been discovered that some of these duality relations can be extended to deformed MICZ-Kepler systems and 4D singular oscillators that are sums of two 2D singular oscillators. It was proven that the superintegrability and multiseparability properties are preserved \cite{Mard5,Mard6,Ran1,Sal1,Tre1,Pet1}. The dual of the 4D singular oscillator has interesting properties and its classical analog has period motion \cite{Tre1}. It has been recognized this is also the case for the dual of the 8D singular oscillator \cite{Mar1}. Moreover, it has been shown that a quadratic algebra structure exists for these two models with monopole interactions and their duals \cite{Mar1,Mar2}. \par
%
%
In this paper, we introduce a new family of $N$-dimensional superintegrable singular oscillators with arbitrary partition ($n$, $N-n$) of the coordinates. This model is a generalization of the 4D and 8D systems \cite{Pet1,Mar1, Mar2} obtained from monopole systems via the Hurwitz transformations. However, even in 4 and 8 dimensions only the symmetric cases $(2,2)$ and $(4,4)$ were studied \cite{Mard6,Mar2}. We show that quadratic algebra structures exist for all members of the family and can be used to obtain the energy spectrum. Another main objective of this paper is to extend the analysis presented in \cite{FH1} to more complicated algebraic structures. \par
%
%
The paper is organized as follow. In Section 2, we present the integrals of motion of the new family of superintegrable systems. In Section 3, we obtain the quadratic algebra and present the Poisson analog and their Casimir operator. We also highlight the structure of higher rank quadratic algebra and the decomposition. In Section 4, we generalize the realizations as deformed oscillator algebra, construct the Fock space and obtain the finite-dimensional unitary representations (unirreps). We also using this analysis provide an algebraic derivation of the energy spectrum. In Section 5, we use the method of separation of variables in double hyperspherical coordinates and compare with the results obtain algebraically. In the closing section 6, we present some discussion with few remarks on the physical and mathematical relevance of these algebras. 

\section{New family of superintegrable system}
Let us consider a family of $N$-dimensional superintegrable Hamiltonians involving singular terms with any two partitions of the coordinates $(n, N-n)$
\begin{eqnarray}
H=\frac{p^2}{2}+\frac{\omega^2 r^2}{2}+\frac{c_1}{x^2_1+...+x^2_n}+\frac{c_2}{x^2_{n+1}+...+x^2_N},\label{hamil}
\end{eqnarray}
where $ \vec{r}=(x_{1},x_{2},...,x_{N})$, $\vec{p}=(p_{1},p_{2},...,p_{N})$, $r^{2}=\sum_{i=1}^{N}x_{i}^{2}$, $p_{i}=-i \hbar \partial_{i}$ and $c_1$, $c_2$ are positive real constants. This family of systems represents the sum of two singular oscillators of dimensions $n$ and $N-n$. It is a generalization of the 4D and 8D systems obtained via the Hurwitz transformation for specific case (2,2) in 4D \cite{Pet1,Mar2} and (4,4) in 8D \cite{Mar1}. In fact model (\ref{hamil}) not only contains cases (2,2) and (4,4), but also cases (1,3) and (1,7), (2,6), (3,5) for 4D and 8D respectively. An algebraic derivation of the energy spectrum has previously been obtained only for the two symmetric cases (2,2) and (4,4). 

The system (\ref{hamil}) has the following integrals of motion
\begin{eqnarray}
A&=&-\frac{h^2}{4}\left\{\sum^N_{i,j=1}x^2_i\partial^2_{x_j}-\sum^N_{i,j=1}x_i x_j \partial_{x_i}\partial_{x_j}-(N-1)\sum^N_{i=1}x_i \partial_{x_i}\right\}\nonumber\\&& +\frac{1}{2}\sum^N_{i=1}x^2_i\left\{\frac{c_1}{x^2_1+...+x^2_n}+\frac{c_2}{x^2_{n+1}+...+x^2_N}\right\},\label{kpA}
\end{eqnarray}
\begin{eqnarray}
B&=&\frac{1}{2}\left\{\sum^n_{i=1}p^2_i-\sum^N_{i=n+1}p^2_i\right\} + \frac{\omega^2}{2}\left\{\sum^n_{i=1}x^2_i-\sum^N_{i=n+1}x^2_i\right\}\nonumber\\&&+\frac{c_1}{x^2_1+...+x^2_n}+\frac{c_2}{x^2_{n+1}+...+x^2_N},\label{kpB}
\end{eqnarray}
which can be verified to fulfill the commutation relation
\begin{eqnarray*}
[H,A]=0=[H,B].
\end{eqnarray*}
The first order integrals of motion are given by 
\begin{eqnarray}
&&J_{ij}=x_i p_j-x_j p_i,\qquad i, j= 1,2,....,n,
\\&&K_{ij}=x_i p_j-x_j p_i,\qquad i, j= n+1,....,N.
\end{eqnarray}
The integrals of motion $A$ and $B$ are associated with the separation of variables in double hyper-Eulerian and double hyperspherical coordinates respectively.
Let
\begin{eqnarray}
J_{(2)}=\sum_{i<j}J^2_{ij}, \qquad  K_{(2)}=\sum_{i<j}K^2_{ij}\label{kpJ}.
\end{eqnarray}
$J_{(2)}$ and $K_{(2)}$ represent the second order Casimir operators and fulfill the commutation relations
\begin{eqnarray}
[H,J_{(2)}]&=&[H,K_{(2)}]=[A,J_{(2)}]=[A,K_{(2)}]=[B,J_{(2)}]\nonumber\\&&=[B,K_{(2)}]=[J_{(2)},K_{(2)}]=0.
\end{eqnarray}
These commutation relations can be conveniently represented by the following diagrams
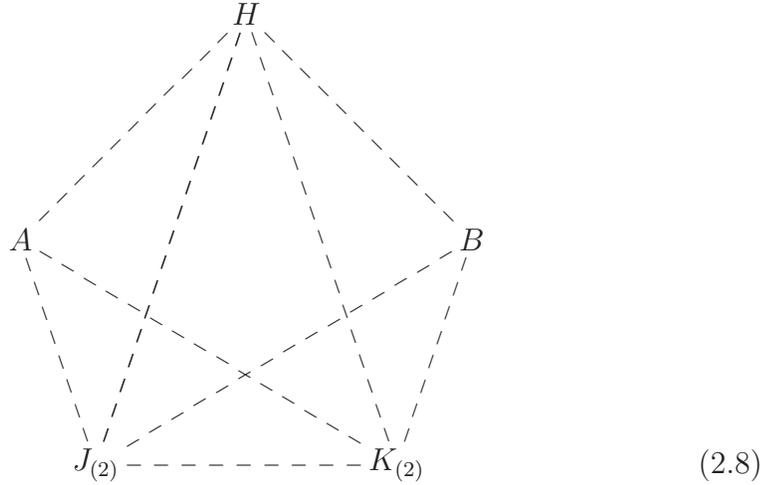
\begin{equation}
\begin{xy}
(10,0)*+{J_{(2)}}="f"; (50,0)*+{K_{(2)}}="k"; (0,30)*+{A}="a"; (60,30)*+{B}="b"; (30,60)*+{H}="h";  
"f";"k"**\dir{--}; 
"h";"b"**\dir{--};
"h";"f"**\dir{--};
"h";"a"**\dir{--};
"h";"k"**\dir{--};
"h";"f"**\dir{--};
"a";"f"**\dir{--}; 
"a";"k"**\dir{--};
"b";"f"**\dir{--};
"b";"k"**\dir{--};
\end{xy}
\end{equation}

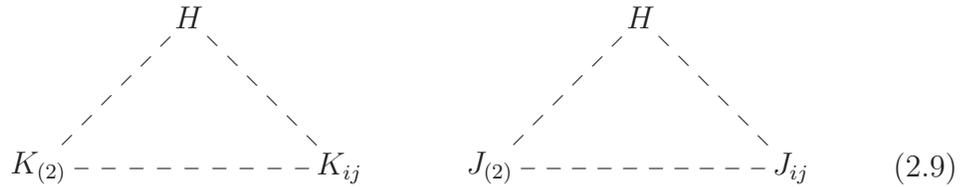
\begin{equation}
\begin{xy}
(0,0)*+{K_{(2)}}="m"; (40,0)*+{K_{ij}}="n"; (20,20)*+{H}="r";   (60,0)*+{J_{(2)}}="j"; (100,0)*+{J_{ij}}="l"; (80,20)*+{H}="g"; 
"j";"l"**\dir{--}; 
"l";"g"**\dir{--};
"g";"j"**\dir{--};
"m";"n"**\dir{--}; 
"n";"r"**\dir{--};
"r";"m"**\dir{--};
\end{xy}
\end{equation}
The first diagram shows that $J_{(2)}$ and $K_{(2)}$ are central elements. The second and third diagrams show that $J_{(2)}$ and $K_{(2)}$ are the Casimir operators of $so(n)$ and $so(N-n)$ Lie algebras realized by angular momentum $J_{ij}$, $ i, j= 1,2,....,n$ and $K_{ij}$, $i, j= n+1,....,N$ respectively. The models are minimally superintegrable and the total number of algebraically independant integrals is $N+1$. The construction of the minimally superintegrable systems is interesting, as research has so far mainly focused on maximally superintegrable systems. In the next section, we construct the quadratic algebra, its Casimir operator and finite dimensional unitary representations which give the energy spectrum of the superintegrable systems.

\section{Quadratic algebra structure}

We present in this section quadratic algebra structure $Q(3)$ of the superintegrable Hamiltonian systems (\ref{hamil}). We show how the $su(N)$ symmetry algebra of isotropic harmonic oscillator is broken to $Q(3)\oplus so(n)\oplus so(N-n).$ 

\subsection{The isotropic harmonic oscillator and $su(N)$}
In this subsection we review some facts about isotropic harmonic oscillator and $su(N)$. The Hamiltonian (\ref{hamil}) in the limit $c_{1}=0$ and $c_{2}=0$ reduces to the isotropic harmonic oscillator in $N$-dimension is given by 
\begin{eqnarray}
H=\frac{p^2}{2}+\frac{\omega^2 r^2}{2}.\label{Iohamil}
\end{eqnarray}
In this limiting case, we define the following operators \cite{Budin1} by 
\begin{eqnarray}
&&a_i=\frac{1}{\sqrt{2}}(p_i+i\omega x_i),\quad a^+_i=\frac{1}{\sqrt{2}}(p_i-i\omega x_i),
\end{eqnarray}
which satisfy the commutation relation $[a_i,a^+_j]=-\omega\hbar\delta_{ij}$. Furthermore, we can define the operators 
\begin{eqnarray}
F^j_i=\frac{1}{\sqrt{2\omega\hbar}}\{a_i, a^+_j\}
\end{eqnarray}
are hermitian and satisfy the commutation relations
\begin{eqnarray}
[F^i_j, F^k_l]=\delta_{il} F^k_j-\delta_{jk} F^i_l.
\end{eqnarray}
Hence $su(N)$ is the symmetry algebra of the isotropic harmonic oscillator.
We present some key briefly key elements related to an algebraic derivation of the energy spectrum. The $su(N)$ symmetry algebra can be embedded in the non-compact algebra $sp(N)$, define the operators
\begin{eqnarray}
F^{ij}_0=\frac{-1}{\sqrt{2\omega\hbar}}\{a^+_i, a^+_j\}, \quad F_{ij}^0=\frac{1}{\sqrt{2\omega\hbar}}\{a_i, a_j\}
\end{eqnarray}
and satisfying the commutators
\begin{eqnarray}
[F^i_j, F^{lk}_0]=\delta_{jk} F^{il}_0+\delta_{jl} F^{ik}_0,
\quad
[F^i_j, F_{lk}^0]=-\delta_{ik} F_{jl}^0-\delta_{il} F_{jk}^0.
\end{eqnarray}
One can construct the Casimir operator of $sp(N)$ in the form
\begin{eqnarray}
Q_2=B_2+\frac{H^2}{2}+\frac{1}{\sqrt{2\omega\hbar}}\{F_{ij}^0,F^{ij}_0\},
\end{eqnarray}
where $B_2$ is the Casimir operator of $su(N)$.
The eigenvalues of $Q_2$ and $B_2$ in the representations of $sp(N)$ and $su(N)$ are as 
\begin{eqnarray}
Q_2=-\frac{N}{2}(N+\frac{1}{2}), \quad B_2=-\frac{l(l+N)(N-1)}{N},
\end{eqnarray}
where all representations belonging to the $H$-eigenstates are obtained for $l=1,2,\dots ,$
and the energy spectrum
\begin{eqnarray}
E=\omega \hbar (l+\frac{N}{2}),\quad l=0,1,2,\dots
\end{eqnarray}

\subsection{Quadratic Poisson algebra }
We study in this subsection the Poisson algebra of the classical version of the superintegrable system (\ref{hamil}) and its Casimir operators. The system has the second order integrals of motion, 
\begin{eqnarray}
A&=&\frac{1}{4}\left\{\sum^N_{i,j=1}x^2_ip^2_j-\sum^N_{i,j=1}x_i x_j p_i p_j\right\}\nonumber\\&& +\frac{1}{2}\sum^N_{i=1}x^2_i\left\{\frac{c_1}{x^2_1+...+x^2_n}+\frac{c_2}{x^2_{n+1}+...+x^2_N}\right\},
\end{eqnarray}
$B$ and $J_{(2)}$, $K_{(2)}$ are given by (\ref{kpB}) and (\ref{kpJ}) respectively. 
It can be verified they satisfy 
\begin{eqnarray}
\{H,A\}_{p}&=&\{H,B\}_p=0=\{H,J_{(2)}\}_p=\{H,K_{(2)}\}_p,
\end{eqnarray}
where $\{,\}_{p}$ is the usual Poisson bracket. Also we can check these following Poisson brackets
\begin{eqnarray*}
\{A,J_{(2)}\}_p=\{A,K_{(2)}\}_p=\{B,J_{(2)}\}_p=0=\{B,K_{(2)}\}_p=\{J_{(2)},K_{(2)}\}_p.
\end{eqnarray*}
The above commutation relations show that $J_{(2)}$ and $K_{(2)}$ are second order Casimir operators and central elements. We now construct a new integral of motion from $A$ and $B$ as 
\begin{eqnarray}
\{A,B\}_p=C.
\end{eqnarray} 
The integral of motion $C$ is cubic function of momenta. After a direct but involving computation relying on properties of the Poisson bracket and identities, we can show that the integrals of motion generate the quadratic Poisson algebra $QP(3)$, 
\begin{eqnarray}
&&\{A,B\}_p=C,
\\&&\{A,C\}_p= -4AB+ J_{(2)}H- K_{(2)}H+2(c_1-c_2)H,
\\&&\{B,C\}_p=2 B^2-2 H^2+16\omega^2 A-4\omega^2 J_{(2)}-4\omega^2 K_{(2)}-8\omega^2 (c_1+c_2).\nonumber\\&&
\end{eqnarray}
The Casimir operator of this quadratic Poisson algebra can be shown to be cubic and explicitly given by 
\begin{eqnarray}
K&=&C^2+4AB^2-2[J_{(2)}H-K_{(2)}H+2(c_1-c_2)H]B+16\omega^2 A^2\nonumber\\&&-2[8\omega^2 (c_1+c_2)+4\omega^2 J_{(2)}+4\omega^2 K_{(2)}+2H^2]A.
\end{eqnarray}
By means of explicit expressions as functions of the coordinates and the momenta for the generators $A$, $B$, $C$ and the central elements, the Casimir operator becomes
\begin{eqnarray*}
K_1&=&-2 J_{(2)}H^2-2 K_{(2)}H^2-4(c_1+c_2)H^2-\omega^2 J^2_{(2)}-\omega^2 K^2_{(2)}+2\omega^2 J_{(2)} K_{(2)}\\&&-4\omega^2(c_1-c_2)J_{(2)}+4\omega^2(c_1-c_2)K_{(2)}-4\omega^2(c_1-c_2)^2.
\end{eqnarray*}
The quadratic Poisson algebra and the Casimir operator are useful in deriving the quadratic algebra and Casimir operator: the lowest order terms of $\hbar$ in quantum case coincide with Poisson analog. The first order integrals of motion generate a Poisson algebra isomorphic to $so(n)$ Lie algebra
\begin{eqnarray*}
&\{J_{ij},J_{kl}\}_{p}&= \delta_{ik}J_{jl}+ \delta_{jl}J_{ik}-\delta_{il}J_{jk}-\delta_{jk}J_{il}, 
\end{eqnarray*}
for $i, j, k, l=1, ..., n$ and $so(N-n)$ Lie algebra
\begin{eqnarray}
&\{K_{ij},K_{kl}\}_{p}&= \delta_{ik}K_{jl}+ \delta_{jl}K_{ik}-\delta_{il}K_{jk}-\delta_{jk}K_{il},
\end{eqnarray}
for $i, j, k, l=n+1, ..., N-n$.
So the full symmetry algebra is a direct sum of the quadratic Poisson algebra $QP(3)$, $so(n)$ and $so(N-n)$ Lie algebras.

\subsection{Quadratic algebra }
We now construct integral of motion $C$ of the quantum system from (\ref{kpA}) and (\ref{kpB}) via commutator
\begin{eqnarray}
[A,B]=C, \label{kpC}
\end{eqnarray}
where $C$ represents a new integral of motion and is a cubic function of momenta. The cubicness of $C$ explains the impossibility of expressing $C$ as a polynomial function of other integrals of motion, which are quadratic function of momenta. After an involving but direct computations that is based on properties of commutators and various identities, we obtain the following quadratic algebra $Q(3)$ of the integrals of motion $H$, $A$ and $B$ 
\begin{eqnarray}
[A,C]&=& 2\hbar^2 \{A,B\}-\hbar^2 J_{(2)}H+\hbar^2 K_{(2)}H-\frac{\hbar^2}{4} \left\{8c_1-8c_2\right.\nonumber\\&&\left.-(N-4)(N-2n)\hbar^2\right\}H+\frac{\hbar^4}{4}N(N-4)B,\label{kpAC}
\end{eqnarray}
\begin{eqnarray}
[B,C]&=&-2\hbar^2 B^2+2\hbar^2 H^2-16\hbar^2\omega^2A+4\hbar^2\omega^2 J_{(2)}+4\hbar^2\omega^2 K_{(2)}\nonumber\\&&+8\hbar^2\omega^2\{c_1+c_2-\frac{\hbar^2}{4}n(N-n)\}.\label{kpBC}
\end{eqnarray}
It can be demonstrated, the Casimir operator is a cubic expression of the generators and is explicitly given in terms of the generators ($A,B$ and $C$) as
\begin{eqnarray}
K&=&C^2-2\hbar^2\{A,B^2\}+\frac{\hbar^4}{4}\{16-N(N-4)\}B^2+2\hbar^2\left[J_{(2)}H-K_{(2)}H\right.\nonumber\\&&\left.+\frac{1}{4}\{8c_1-8c_2-(N-4)(N-2n)\hbar^2\}H\right]B-16\hbar^2\omega^2 A^2\nonumber\\&&+2\hbar^2\left[8\omega^2\{c_1+c_2-\frac{\hbar^2}{4}n(N-n)\}+4\omega^2 J_{(2)}+4\omega^2 K_{(2)}+2H^2\right]A.\nonumber\\&&\label{kpK}
\end{eqnarray}
Using the realization of the integrals of motion $A$, $B$, $C$ and the central element as differential operators, we can represent the Casimir operator (\ref{kpK}) only in terms of the central elements $H$, $J_{(2)}$ and $K_{(2)}$, 
\begin{eqnarray}
K&=&2\hbar^2 J_{(2)}H^2+2\hbar^2 K_{(2)}H^2+\frac{\hbar^2}{4}\left[16c_1+16c_2-\{4(N-4)\right.\nonumber\\&&\left.-(N-2n)^2\}\hbar^2\right]H^2+\hbar^2\omega^2 J^2_{(2)}+\hbar^2\omega^2 K^2_{(2)}-2\hbar^2\omega^2 J_{(2)} K_{(2)}\nonumber\\&&+4\hbar^2\omega^2\{c_1-c_2-\frac{1}{4}(N-4)(N-n)\hbar^2\}J_{(2)}-4\hbar^2\omega^2\{c_1-c_2\nonumber\\&&+\frac{1}{4}n(N-4)\hbar^2\}K_{(2)}+4\hbar^2\omega^2\left[(c_1-c_2)^2-\frac{1}{2}(N-n)(N-4)\hbar^2 c_1\right.\nonumber\\&&\left.-\frac{1}{2}n(N-4)\hbar^2 c_2+\frac{1}{4}n(N-n)(N-4)\hbar^4\right].\label{kpK1}
\end{eqnarray}
This is key step in the application of the deformed oscillator algebra approach which relies on both form of the Casimir operators. The first order integrals of motion, that are simply components of angular momentum, generate an algebra isomorphic to the $so(n)$ Lie algebra
\begin{eqnarray*}
&[J_{ij},J_{kl}]&= i(\delta_{ik}J_{jl}+ \delta_{jl}J_{ik}-\delta_{il}J_{jk}-\delta_{jk}J_{il})\hbar, 
\end{eqnarray*}
for $i, j, k, l=1, ..., n$ and  $so(N-n)$ Lie algebra
\begin{eqnarray*}
&[K_{ij},K_{kl}]&=i( \delta_{ik}K_{jl}+ \delta_{jl}K_{ik}-\delta_{il}K_{jk}-\delta_{jk}K_{il})\hbar,
\end{eqnarray*}
for $i, j, k, l=n+1, ..., N-n$. So the full symmetry algebra is a direct sum of the quadratic algebra $Q(3)$, $so(n)$ and $so(N-n)$ Lie algebras. Thus the $su(N)$ Lie algebra generated by the integrals of motion of the $N$-dimensional isotropic harmonic oscillators is deformed into higher rank quadratic algebra $Q(3) \oplus so(n) \oplus so(N-n)$ for the family of superintegrable systems (\ref{hamil}). The structure constants of the quadratic algebra involve three central elements, the Hamiltonian and the two Casimir operators of the Lie algebras approached in the decomposition.

\section{Energy spectrum and unirreps}

We now consider the realizations of the quadratic algebra ((\ref{kpC})-(\ref{kpBC})) in terms of deformed oscillator algebra \cite{Das1,Das2} $\{\aleph, b^{\dagger}, b\}$ defined by
\begin{eqnarray}
[\aleph,b^{\dagger}]=b^{\dagger},\quad [\aleph,b]=-b,\quad bb^{\dagger}=\Phi (\aleph+1),\quad b^{\dagger} b=\Phi(\aleph),
\end{eqnarray}
where $\aleph $ is the number operator and the function $\Phi(x)$ is well behaved real function satisfying the boundary condition
\begin{eqnarray}
\Phi(0)=0, \quad \Phi(x)>0, \quad \forall x>0.\label{kpbc}
\end{eqnarray}
The $\Phi(x)$ is the so-called structure function. The realization of $Q(3)$ is of the form $A=A(\aleph)$, $B=b(\aleph)+b^{\dagger} \rho(\aleph)+\rho (\aleph)b$, where $A(x)$, $b(x)$ and $\rho(x)$ are functions. Similar to the quadratic algebra for 2D superintegrable systems \cite{Das1}, we have
\begin{eqnarray}
&&A(\aleph)=\hbar^2\{(\aleph+u)^2-\frac{(N-2)^2}{16}\},
\\&&
B(\aleph)=\frac{8c_1-8c_2+4J_{(2)}-4K_{(2)}+(4N-8n+2nN-N^2)\hbar^2}{16\hbar^2\{(\aleph+u)^2-\frac{1}{4}\}},
\\&&
\rho(\aleph)=\frac{1}{3.2^{20}.\hbar^{16}(\aleph+u)(1+\aleph+u)\{1+2(\aleph+u)\}^2},
\end{eqnarray}
where $u$ is a constant ( in fact a representation dependent constant ) to be determined from the constraints on the structure function. We construct the structure function $\Phi(x)$ by using the realization, the quadratic algebra ((\ref{kpC}), (\ref{kpAC}), (\ref{kpBC})) and the two forms of the Casimir operator (\ref{kpK}) and (\ref{kpK1})
\begin{eqnarray}
\Phi(x;u,H)&=&12288 \hbar^{12} \left[64 c_1^2 + 64 c_2^2 - 48 \hbar^4 - 32 \hbar^2 J_{(2)} + 16 J_{(2)}^2 -32\hbar^2 K_{(2)}\nonumber \right.\\&&\left.- 32 J_{(2)} K_{(2)} + 16 K_{(2)}^2 - 64 \hbar^2 J_{(2)} n + 64 \hbar^2 K_{(2)} n + 48 \hbar^4 n^2 \nonumber \right.\\&&\left.+ 32 \hbar^4 N + 32 \hbar^2 J_{(2)} N- 32 \hbar^2 K_{(2)} N - 48 \hbar^4 n N + 16 \hbar^2 J_{(2)} n N\nonumber \right.\\&&\left. - 16 \hbar^2 K_{(2)} n N - 32 \hbar^4 n^2 N + 8 \hbar^4 N^2 - 8 \hbar^2 J_{(2)} N^2 + 8 \hbar^2 K_2 N^2 \nonumber\right.\\&&\left.+ 32 \hbar^4 n N^2 + 4 \hbar^4 n^2 N^2 - 8 \hbar^4 N^3 - 4 \hbar^4 n N^3 + \hbar^4 N^4 \nonumber\right.\\&&\left.-16 c_2 [4 (J_{(2)} - K_{(2)}) + \hbar^2 \{(N-4)(2n-N)+ 4 (1 - 2(x+u))^2\}]\nonumber \right.\\&&\left.- 16 c_1 [8 c_2 - 4 J_{(2)} + 4 K_{(2)} + \hbar^2\{(N-4)(N-2n) \nonumber \right.\\&&\left.+ 4 (1 - 2(x+u))^2\}]+ 32 \hbar^2 [4 (J_{(2)} + K_{(2)}) + \hbar^2 \{2 n^2 + (N-2)^2 \nonumber \right.\\&&\left.- 2 n N\}] (x+u) - 32 \hbar^2 [4 (J_{(2)}+ K_{(2)}) + \hbar^2 \{2 (n^2-2)\nonumber \right.\\&&\left.- 2 (n+2) N + N^2\}] (x+u)^2 - 512 \hbar^4 (x+u)^3 + 256 \hbar^4 (x+u)^4\right]\nonumber \\&&\times [H^2 - h^2 \{1 - 2 (x+u)\}^2 \omega^2].\label{kpST}
\end{eqnarray}
We will show how the finite dimensional unirreps can be obtained using an appropriate Fock space. In 2D, we need $|n, E>$ such that $\aleph|n, E>=n|n, E>$. However, in our case one needs to define quantum numbers associated to certain subalgebra chain. We use two subalgebra chains, $so(n) \supset so(n-1) \supset ... \supset so(2)$  and $so(N-n) \supset so(N-n-1) \supset ... \supset so(2)$ related chains of quadratic Casimir operators \cite{Ras1} $J_{2}^{(\alpha)}$ and $K_{2}^{(\alpha)}$ can be written as
\begin{eqnarray}
&&J_{2}^{(\alpha)}=\sum^{\alpha}J_{ij},\quad \alpha=2,...,n,
\\&&
K_{2}^{(\alpha)}=\sum^{\alpha}K_{ij},\quad \alpha=n+2,...,N.
\end{eqnarray}
In fact, $|n, E>$ means $|n,E,l_{N},...,l_{n+2},l_{n},...,l_{2}>$. Then the eigenvalues of $J_{(2)}$ and $K_{(2)}$ are $\hbar^2 l_{n}(l_{n}+n-2)$ and $\hbar^2 l_{N-n}(l_{N-n}+N-n-2)$ respectively. The constraint (\ref{kpbc}) can be imposed on Fock type representation of the deformed oscillator algebra, $|n, E>$ with $\aleph|n, E>=n|n, E>$,  and $H$ in $\Phi(x, u, H)$ replaced by $E$.  Hence by using the eigenvalues of  $J_{(2)}$, $K_{(2)}$ and $H$, the structure function becomes the following factorized form:
\begin{eqnarray}
\Phi(x)&=&-12582912 \hbar^{18}\omega^2 [x+u-\frac{1}{4}(2+m_1+m_2)] [x+u-\frac{1}{4}(2-m_1+m_2)]\nonumber\\&&\times[x+u-\frac{1}{4}(2+m_1-m_2)] [x+u-\frac{1}{4}(2-m_1-m_2)]\nonumber\\&&\times(x+u -\frac{-H +\hbar \omega}{2\hbar \omega})(x - \frac{H + \hbar \omega}{2 \hbar \omega}),\label{pro3}
\end{eqnarray}
where
$\hbar^2 m_1^2 = 8 c_1  + 4 J_{(2)} +\hbar^2 (n-2)^2$ and 
$\hbar^2 m_2^2 = 8 c_2  + 4 K_{(2)} + \hbar^2(N-n-2)^2$.
For unirreps to be finite dimensional, we impose the following constrains on the structure function:
\begin{equation}
\Phi(p+1; u, E)=0;\quad \Phi(0;u,E)=0;\quad \Phi(x)>0,\quad \forall x>0,\label{pro2}
\end{equation}
where $p$ is a positive integer and $u$ is arbitrary constant. We then obtain finite $(p+1)$-dimensional unirreps. The solution of the constraints (\ref{pro2}) gives the energy $E$ and constant $u$. Thus we obtain the following possible structure functions and energy spectra, for $\epsilon_1=\pm 1$, $\epsilon_2= \pm 1$, $\eta =24576 \hbar^{18}\omega^2$.
\\Set-1:
\begin{eqnarray*}
u = \frac{-E + \hbar \omega}{2\hbar \omega},\qquad E = 2\hbar\omega(p+1+\frac{\epsilon_1 m_1 +\epsilon_2  m_2 }{4} ),
\end{eqnarray*}
\begin{eqnarray*} 
\Phi(x) &=& \eta x \hbar^{18}\omega^2 [4+4p-4x-(1-\epsilon_1) m_1 +(1+\epsilon_2)m_2]\\&&\times[4+4p-4x+(1+\epsilon_1) m_1 -(1-\epsilon_2)m_2][4+4p-4x\nonumber\\&&-(1-\epsilon_1) m_1 -(1-\epsilon_2)m_2][4+4p-4x+(1+\epsilon_1) m_1 \nonumber\\&&+(1+\epsilon_2)m_2][4+4p-2x+ \epsilon_1 m_1 +\epsilon_2 m_2].
\end{eqnarray*}
\\Set-2:
\begin{eqnarray*}
u = \frac{E + \hbar \omega}{2\hbar \omega},\qquad E = 2\hbar\omega(p+1+\frac{\epsilon_1 m_1 +\epsilon_2  m_2 }{4} ),
\end{eqnarray*}
\begin{eqnarray*} 
\Phi(x) &=& - x [4+4p+4x-(1-\epsilon_1) m_1 +(1+\epsilon_2)m_2]\\&&\times[4+4p+4x+(1+\epsilon_1) m_1 -(1-\epsilon_2)m_2][4+4p+4x\nonumber\\&&-(1-\epsilon_1) m_1 -(1-\epsilon_2)m_2][4+4p+4x+(1+\epsilon_1) m_1 \nonumber\\&&+(1+\epsilon_2)m_2][4+4p+2x+ \epsilon_1 m_1 +\epsilon_2 m_2].
\end{eqnarray*}
\\Set-3:
\begin{eqnarray*}
u = \frac{1}{4}(2+\epsilon_1 m_1+\epsilon_2 m_2),\qquad E = 2\hbar\omega(p+1+\frac{\epsilon_1 m_1 +\epsilon_2  m_2 }{4} ),
\end{eqnarray*}
\begin{eqnarray*} 
\Phi(x)& =& \eta (p+1-x)[4x-(1-\epsilon_1) m_1 -(1-\epsilon_2)m_2]\\&&\times[4x+(1+\epsilon_1) m_1 -(1-\epsilon_2)m_2][4x-(1-\epsilon_1) m_1 +(1+\epsilon_2)m_2]\\&&\times[4x+(1+\epsilon_1) m_1 +(1+\epsilon_2)m_2][2+2p+2x+ \epsilon_1 m_1 +\epsilon_2 m_2].
\end{eqnarray*}
The structure functions $\Phi(x)>0$ for $\varepsilon_{1}=1$, $\varepsilon_{2}=1$ and $m_{1}, m_{2}>0$. In the limit $c_1=0$ and $c_2=0$, this results coincide with $N$-dimensional harmonic oscillator with the following relation among the quantum numbers $l=2p+l_n+l_{N-n}$ and the algebraic derivation using the $su(N)$ and $sp(N)$ Lie algebra and their Casimir operators and eigenvalues. Let us mention that the value of $\eta$ do not play a role, only the sign needs to be taken into account for the constraint to obtain the finite- dimensional unirreps.

\section{Separation of variables }  
Each member of the family of superintegrable Hamiltonian systems (\ref{hamil}) is multiseparable and allows for the Schordinger equation the separation of variables in double hyper Eulerian and double hyperspherical coordinates. We can rewrite (\ref{hamil}) into the sum of two singular oscillators of dimensions $n$ and $N-n$ as 
\begin{eqnarray}
H=H_1+H_2,
\end{eqnarray}
where
 \begin{eqnarray}
 &&H_1=\frac{1}{2}(p^2_1+...+p^2_n)+\frac{\omega^2 r^2_1}{2}+\frac{c_1}{r^2_1},
 \\&&
 H_2=\frac{1}{2}(p^2_{n+1}+...+p^2_N)+\frac{\omega^2 r^2_2}{2}+\frac{c_2}{r^2_2}
 \end{eqnarray}
 and the position vectors $r^2_1=x^2_1+...+x^2_n, \quad r^2_2=x^2_{n+1}+...+x^2_N.$
The Schrodinger equation $H\psi(r,\Omega)=E\psi(r, \Omega)$ can also be written as 
\begin{eqnarray}
 H_1\psi(r_1,\Omega)=E_1\psi(r_1, \Omega), \qquad H_2\psi(r_2,\Omega)=E_2\psi(r_2, \Omega),
\end{eqnarray} 
where $E_1$ and $E_2$ are the eigenvalues of $H_1$ and $H_2$ respectively.
The $\mathcal{N}$-dimensional hyperspherical coordinates are given by 
\begin{eqnarray}
&x_{1}&=r \sin(\Phi_{\mathcal{N}-1})\sin(\Phi_{\mathcal{N}-2})\cdots \sin(\Phi_{1}),
\nonumber\\& x_{2}&=r \sin(\Phi_{\mathcal{N}-1})\sin(\Phi_{\mathcal{N}-2})\cdots \cos(\Phi_{1}),
\nonumber\\&...&
\nonumber\\&...&
\nonumber\\&  x_{\mathcal{N}-1}&=r \sin(\Phi_{\mathcal{N}-1})\cos(\Phi_{\mathcal{N}-2}),
\nonumber\\&  x_{\mathcal{N}}&=r \cos(\Phi_{\mathcal{N}-1}),
\end{eqnarray}
where the $\mathcal{N}$ $x_{i}$'s are Cartesian coordinates in the hyperspherical coordinates, $\{\Phi_{1},\dots, \Phi_{\mathcal{N}-1}\}$ are the hyperspherical angles and $r$ is the hyperradius. We can also introduce a double type of hyperspherical coordinates system by considering two copies and setting $\mathcal{N}=n$ and $\mathcal{N}=N-n$ respectively for the $H_{1}$ component and the $H_{2}$ component. The Schrodinger equation of $H_1$ in $n$-dimensional hyperspherical coordinates 
\begin{eqnarray}
\left[\frac{\partial^2}{\partial r^2_1}+\frac{n-1}{r_1}\frac{\partial}{\partial r_1}-\frac{1}{r^2_1}\Lambda^2(n)-\omega'^2 r^2_1-\frac{2c'_1}{r^2_1}+2E'_1\right]\psi(r_1,\Omega)=0,\label{kpH1}
\end{eqnarray}
where $c'_1=\frac{c_1}{\hbar^2}, \quad\omega'=\frac{\omega}{\hbar}, \quad E'_1=\frac{E_1}{\hbar^2}$ and the grand angular momentum operator $\Lambda^2(n)$ which satisfies the recursive formula
\begin{eqnarray}
-\Lambda^2(n)=\frac{\partial^2}{\partial\Phi^2_{n-1}}-(n-2) \cot(\Phi_{n-1})\frac{\partial}{\partial\Phi_{n-1}}-\frac{\Lambda^2(n-1)}{\sin^2(\Phi_{n-1})}, 
\end{eqnarray}
valid for $n>0$ and $\Lambda^2(1)=0$. The separation of radial and angular parts  of (\ref{kpH1}) can be perfomed by setting $\psi(r_1,\Omega)=R_1(r_1)y(\Omega_{n-1})$. Thus, we obtain
\begin{eqnarray}
\frac{\partial^2 R_1(r_1)}{\partial r^2_{1}}+\frac{n-1}{r_1} \frac{\partial R_1(r_1)} {\partial r_1}+\{2E'_1-\omega'^2 r^2_1-\frac{2c'_1}{r^2_1}-\frac{l_{n}(l_{n}+n-2)}{r^2_1}\}R_1(r_1)=0,\nonumber\\\label{KF1}
\end{eqnarray}
\begin{eqnarray}
\Lambda^2(n)y(\Omega_{n-1})=l_{n}(l_{n}+n-2)y(\Omega_{n-1}),
\end{eqnarray}
where $l_{n}(l_{n}+n-2)$ being the general form of the separation constant.
Now the radial equation (\ref{KF1}) can be converted, by setting $u=a r^2_1$, $R_1(u)=u^\alpha f(u)$ and $f(u)=e^{\beta u}f_1(u)$, to
\begin{eqnarray}
u f''_1(u)+\{2(\delta_1+\frac{1}{2}l_{n}+\frac{n}{4})-u\}f'_1(u)+\{\frac{E'_1}{2\omega'}-(\delta_1+\frac{1}{2}l_{n}+\frac{n}{4})\}f_1(u)=0,\nonumber\\&&\label{kp17}
\end{eqnarray}
where
\begin{eqnarray}
\delta_1&=&\left\{\sqrt{(\frac{1}{2}l_{n}+\frac{n-2}{4})^2+\frac{1}{2}c'_1}-\frac{n-2}{4}\right\}-\frac{1}{2}l_{n}.
\end{eqnarray}
Set 
\begin{eqnarray}
N_1=\frac{E'_1}{2\omega'}-\left(\delta_1+\frac{1}{2}l_{n}+\frac{n}{4}\right).\label{kpN1}
\end{eqnarray}
Then the equation (\ref{kp17}) represents the confluent hypergeometric equation and it gives the solution in terms of special function  
\begin{eqnarray}
\psi_{N_1 M_1}(u)&=&\sqrt{\frac{2\Gamma\{N_1+2(\delta_1+\frac{1}{2}l_{n}+\frac{n}{4})\}}{N_1!}}\frac{a e^{-\frac{u}{2}} u^{(\delta_1+\frac{1}{2}l_{n})/2}}{\sqrt{2(\delta_1+\frac{1}{2}l_{n}+\frac{n}{4})}}\nonumber\\&& \times{}_1F_1\left(-N_1; 2\{\delta_1+\frac{1}{2}l_{n}+\frac{n}{4}\}; u\right)\label{kpWf}
\end{eqnarray}
In order the wavefunctions to be square integrable, the parameter $N_1$ needs to be positive. Hence we obtain the discrete energy $E_1$ of $H_1$ from (\ref{kpN1}) as
\begin{eqnarray}
E_1=2\hbar \omega\left(N_1+\frac{\alpha_1}{2}+\frac{1}{2}\right),\label{EKP1}
\end{eqnarray}
where $\alpha_1=2\delta_1+l_{n}+\frac{n-2}{2}$. The wave equation of $H_2$ in (N-n)-dimensional hyperspherical coordinates provides similar solution $\psi_{N_2 M_2}$, replacing $r_1$, $n$, $l_n$, $c'_1$, $\delta_1$  by $r_2$, $N-n$, $l_{N-n}$, $c'_2$, $\delta_2$ respectively in (\ref{kpWf}) and the energy $E_2$, replacing $N_1$, $\alpha_1$ by $N_2$, $\alpha_2$ respectively in (\ref{EKP1}). Hence the energy spectrum of the $N$-dimensional double singular oscillators
\begin{eqnarray}
E=2\hbar \omega\left(p+1+\frac{\alpha_1+\alpha_2}{2}\right),
\end{eqnarray}
where $p=N_1+N_2$ which coincides with the energy expression obtained algebraically. In fact, this multiseparability properties is also shared by its classical analog as the Hamilton-Jacobi can also be separated in the double type coordinates systems.

\section{Conclusion}

In this paper, we have extended the symmetric double singular oscillators in 4D and 8D  to arbitrary dimensions with any partition $(n, N-n)$ of the coordinates. This provide a new family of minimally superintegrable systems. As main results, we construct and obtain its realization as deformed oscillator algebra. We also construct the finite dimensional unitary representations which enable the algebraic derivation of the energy spectrum. This is compared with the results from the separation of variables method. Moreover, the new family may also include duals of deformed higher dimensional Kepler-Coulomb systems involving nonAbelian monopoles \cite{Vhl1,Men1}. 

Our systems are generalisations of one of the four 2D Smorodinsky-Winternitz models \cite{Win1}. Further generalisations are possible. For example,
\begin{eqnarray}
&H&=\frac{p^2}{2}+\frac{\omega^2 r^2}{2}+\frac{c_1}{x^2_1+\dots +x^2_{n_1}}\nonumber\\&&+\frac{c_2}{x^2_{{n_1}+1}+\dots +x^2_{n_2}}+\dots +\frac{c_\lambda}{x^2_{n_{\lambda-1}+1}+\dots +x^2_{n_\lambda}} 
\end{eqnarray}
generalizes the $N$-dimensional Smorodinsky-Winternitz model \cite{Eva1} to the one with any partition ($n_1, n_2,\dots,n_\lambda$) such that $n_1+n_2+\dots +n_\lambda=N$. This model would also be superintegrable but with a more complicated quadratic algebra and embedded structure seen in \cite{Das3,Tan1}.

{\bf Acknowledgements:}
The research of FH was supported by International Postgraduate Research Scholarship and Australian Postgraduate Award. IM was supported by the Australian Research Council through a Discovery Early Career Researcher Award DE 130101067. YZZ was partially supported by the Australian Research Council, Discovery Project DP 140101492.

\end{document}